\documentclass[11pt]{article}

\usepackage{epsfig}
\usepackage{graphicx}

\begin{document}
\thispagestyle{empty}

\begin{flushright}
UOSTP 06101\phantom{abcd}\\
{\tt hep-th/0603080}\phantom{ab}
\end{flushright}
\vspace{.3cm}

\renewcommand{\thefootnote}{\fnsymbol{footnote}}
\centerline{\Large \bf
 Dual of Big-bang and Big-crunch
}

\vskip 1.2cm
\centerline{\bf Dongsu Bak
}

\vskip 10mm
\renewcommand{\thefootnote}{\arabic{footnote}}
\setcounter{footnote}{0}

\centerline{\it
Physics Department, University of Seoul, Seoul 130-743, Korea}

\vskip 8mm
\centerline{({\tt dsbak@mach.uos.ac.kr})}
\vskip 20mm

\baselineskip 18pt

\begin{quote}
Starting from the Janus solution and its gauge theory
dual, we obtain the dual gauge theory description of the
cosmological solution by the procedure of 
double analytic continuation. The coupling is driven either to zero
or to infinity at the big-bang and big-crunch
singularities, which are shown to be related by the S-duality symmetry.
In the dual Yang-Mills theory description, these are non
singular at all as  the coupling goes to zero in the ${\cal N}$=4 Super
Yang-Mills theory. The cosmological singularities simply signal the
failure of the supergravity description of the full type
IIB superstring theory.
\end{quote}
\vskip 1cm
\centerline{\today}
\newpage
\baselineskip 18pt

\def\nn{\nonumber}

\section{Introduction}
The duality between gravity and
gauge theory seems the key to understand the quantum gravity.
Since the original AdS/CFT
correspondence\cite{Maldacena,Gubser,Witten, Aharony},
there have appeared many variants of the
gravity/gauge theory correspondence. However, until now,
there are no concrete examples of the gauge theory dual to
the cosmological evolution of  gravity theory.

As the
cosmology includes big-bang and, possibly, big crunch
singularities, it would be helpful to understand their nature
once we identify a gauge theory
dual of  cosmological evolution.

Our route to a gauge theory dual of cosmological
evolution is rather straightforward. For this, we take
the Janus
solution\cite{Hirano} as
a starting point. It is a
controlled
nonsupersymmetric deformation of
$AdS_5 \times S^5$
keeping $SO(3,2)\times SO(6)$ part of $SO(4,2)\times SO(6)$
global symmetries, which turns out
to be stable\cite{Hirano,Freedman,Papadimitriou,Celi}.
The five dimensional geometry is
$AdS_4$ sliced while keeping $S^5$ part intact. The dilaton
runs through the bulk as function of the slicing
coordinate $\mu$ and the asymptotic AdS boundary at the
ends of the range $\mu$ has two separate parts joined
at their boundaries. The values of dilaton on the two parts
become different
and the dual Yang-Mills (YM) theory has a Janusian nature. Namely the dual YM
theory  has two domains separated by
a codimension one interface, where the YM coupling jumps
from one to the other\cite{Hirano}. More precise version of the Janus dual YM
theory is proposed in Ref.~\cite{Karch}.
The operator ${\cal L}'_0$ dual to the dilaton
is the variant of the ${\cal N}$=4 $SU(N)$ SYM Lagrange
density in which the scalar kinetic term is $X^I \partial^2 X^I/2$
rather than $-\partial X^I \partial X^I/2$.
The action density for the Janus dual is given by
${\cal L}'_0$ multiplied by the inverse coupling squared
which takes different value in each domain.
There  have been further developments in the studies of
 Janus type solutions\cite{Clark, D'Hoker, D'Hoker1, Hirano1}, its related
issues\cite{Papadimitriou, Mala, Sonner} and  other types of dilatonic
deformation\cite{No,Kehagias,Kehagias1,Gubser1,Girardello, Bak,Yee}.

The standard dictionary of AdS/CFT correspondence says
that  the on-shell supergravity action gives the generating
functional of connected graphs of the dual field
theories, which has validity in the large $N$ limit. Of course
the supergravity should be replaced by the full string
theory for the finite $N$ and string coupling.
The generating functional
has to be renormalized in order to get finite correlation
functions. Thus the generating functional is defined with
help of a renormalization group (RG) scale plus the coupling
 at the RG scale.
The bulk direction may be interpreted as a direction of the
RG scale and, at a finite RG scale, the coupling varies
spatially in a smooth manner.

We note then the time dependent branch of solution
preserving $SO(4,1)\times SO(6)$ symmetries
can be obtained by a specified procedure of double
analytic continuation\footnote{It was recently 
noted in Ref.~\cite{Skenderis1} that
there is a general correspondence between domain wall and 
cosmological solutions.}. The supergravity
solution then obtained is a  Friedmann-Robertson-Walker (FRW)
type cosmology
exhibiting the big-bang and big-crunch singularities.
We define the field theory dual for the cosmological
solution by the procedure of double analytic continuation
of the renormalized generating functional.
If one has full nonperturbative  knowledge including all order
$1/N$ corrections of the Janusian gauge theory, the procedure of
the double analytic continuation can be made precise, leading, in principle,
to the
exact field theory dual of the cosmology.

We carry out the double analytic continuation of the generating functional
in its leading order. The big-bang and big-crunch turns out to be
related by S-duality of IIB string theory and these singularities
arise  because the coupling goes either to zero or to infinity
in the gravity description. We will see, however, that this is simply a
failure of the supergravity description and nothing singular happens
in the ${\cal N}$=4 SYM theory in its weak coupling limit.

The organization of our paper is as follows. In section 2, we will
review the Janus solution and its gauge theory dual.
In section 3, we will
discuss the meaning of the bulk geometry of the Janus solution
 in relation with renormalization.
In section
4, we  will obtain  the cosmological  solution by the procedure
of the double analytic continuation.  In section
5, we will
present the dual gauge theory interpretation of the cosmological solution
including the big-bang and big-crunch singularities.
 We will end with a
discussion of  the other conformal compactification, where the scale change
 becomes the time evolution after the double analytic continuation.

\section{Janus solution and its CFT dual}

The Janus solution\cite{Hirano} corresponds the stable
nonsupersymmetric dilatonic
deformation of $AdS_5 \times S^5$ which preserves
$SO(3,2) \times SO(6)$ part  of the original $SO(4,2) \times SO(6)$
global symmetries. The dilaton is tuned on additionally,
while the original five-form field strength is modified
minimally. 

Following Ref.~\cite{Hirano},
let us review the set of
coordinate systems for the $AdS_d$ space, which may be useful
for our application.
The $AdS_d$ space is defined by a hyperboloid in $R^{2,d-1}$
\begin{equation}
-X_0^2-X_d^2 + X_1^2 +\cdots +X_{d-1}^2 =-1\ .
\label{adsone}
\end{equation}
The global coordinate covers the entire region of the $AdS_d$
space. Using 
the 
parameterization,
\begin{equation}
X_0= {\cos \tau\over \cos \theta},\quad  X_d =
  {\sin\tau  \over \cos\theta}, \quad X_i = \tan\theta\,\, n_i ,\;\;
  i=1,\cdots , d-1\ ,
\label{adstwo}
\end{equation}
with the unit vector  $n_i$
in $R^{d-1}$, the metric on the global
$AdS_d$ is given by
\begin{equation}
ds^2_{AdS_d} = {1\over \cos^2\theta} \big( -d\tau^2 +
  d\theta^2 +\sin^2\theta\, d\Omega_{d-2}^2\big)\ ,
\label{adsthree}
\end{equation}
where $\theta \in \left[0,{\pi/ 2}\right]$.

The Poincar\'e patch is another standard 
parameterization of AdS space, which is dual to the ${\cal N}$=4 SYM
theory on $R^{1,3}$ for the case of $AdS_5 \times S^5$.
By parameterization, $X_0={1 \over
  2}\left(z+\left(1+\vec{x}^2-t^2\right)/z\right)$, $X_d=t/z$,
$X_{i=1,\cdots,d-2}=x_i/z$, and $X_{d-1}={1 \over
  2}\left(z-\left(1-\vec{x}^2+t^2\right)/z\right)$, the metric takes
the form
\begin{equation}
ds_{AdS_d}^2={1\over z^2}\left(-dt^2+d\vec{x}^2+dz^2\right)\ ,
\label{adsnine}
\end{equation}
where $\vec{x}=(x_1,\cdots,x_{d-2})$ and $z\in [0,\infty]$.

For the description of the Janus solution, $AdS_{d-1}$ slicing of
the $AdS_d$ space is very useful.
We use 
$X_{d-1}=w$ as one coordinate ranged over
$[-\infty,+\infty]$
 and  any
coordinate system of the $AdS_{d-1}$ space of radius
$\sqrt{1+w^2}$ for the rest. 
The metric takes then the following form
\begin{equation}
ds_{AdS_d}^2= {dw^2\over 1+w^2} + (1+w^2)
  ds^2_{AdS_{d-1}}\ .
\label{adsfour}
\end{equation}
Defining a new coordinate $\mu$ by $w=\tan\mu$,
the metric is rewritten as
\begin{equation}
ds_{AdS_d}^2= f_0(\mu) \big(d\mu ^2 + ds^2_{AdS_{d-1}
  }\big)\ ,
\label{adsfive}
\end{equation}
where $f_0 (\mu)=1/\cos^2\mu$ and  $\mu \in \left[-{\pi/ 2},
{\pi/2}\right]$.

The conformal boundary of the global AdS metric (\ref{adsthree})
is located at $\theta =\pi/2$ and has the shape of $R\times S^3$
where $R$ is the time direction, while that of the Poincar\'e patch
(\ref{adsnine}) is at $z=0$ and of the shape $R^{1,3}$. In the
$AdS_{d-1}$ slicing
(\ref{adsfive}),
 the appearance of the boundary is less trivial although there is no change of its
 shape.

First consider the case where
the global coordinate is used for the $AdS_{d-1}$ slice in
(\ref{adsfive}). Note then  the metric can be written as
\begin{equation}
ds^2_{AdS_d} = {1\over \cos^2\mu\cos^2\lambda}
 \big( -d\tau^2 + \cos^2\lambda d\mu^2 +
  d\lambda^2 +\sin^2\lambda d\Omega_{d-3}^2\big)\ ,
\label{adssix}
\end{equation}
with $\lambda \in \left[0,{\pi/2}\right]$. The constant time
section of this metric is conformal to a half of $S^{d-1}$, since the
range of $\mu$ is
just from $-{\pi/2}$ to ${\pi/ 2}$. If $\mu$ had  ranged
over $[-\pi,\pi]$, it would have been the full sphere.
 The boundary consists of
two parts, one of which is at $\mu=-\pi/2$ and the other at $\mu=\pi/2$.
These two parts are joined through a surface,
$\lambda=\pi/2$, which is a codimension one interface in the boundary.
The $\mu=\pi/2$ (or  $\mu=-\pi/2$ ) part is 
a half of
$S^{d-2}$ as $\lambda$ ranged over $\left[0,{\pi/ 2}\right]$,
so over all the boundary makes up the full $S^{d-2}$.
For the Janus solution,
the range of $\mu$ becomes
elongated,  but the structure of the conformal boundary remains to be
the same $S^{d-2}$.

Next one may take the Poincar\'e patch for the $AdS_{d-1}$
slice in
(\ref{adsfive}). Then the metric takes the form
\begin{equation}
ds^2_{AdS_d} = {1\over y^2\cos^2\mu}
 \big( -dt^2 + d\vec{x}_{d-3}^2 +dy^2 +y^2 d\mu^2\big)\ ,
\label{adsseven}
\end{equation}
with $y \in [0,\infty]$.  One can easily see that,
by the change of coordinate $x=y\sin\mu$ and $z=y\cos\mu$,
the above metric turns into the conventional form of
the Poincar\'e patch AdS,
\begin{equation}
ds^2_{AdS_d} = {1\over z^2}
 \big( -dt^2 + d\vec{x}_{d-3}^2 +dx^2 + dz^2\big)\ .
\label{adseight}
\end{equation}
Again the boundary consists of two parts, one of which is at $\mu=\pi/2$
and the other at $\mu=-\pi/2$. These two parts are joined through a
codimension one surface, $y=0$, forming a $d-2$ dimensional flat
Euclidean space,
 or $d-1$ dimensional Minkowski space when including the time.
For the Janus solution,
the structure of the conformal boundary remains again the same, on which the
dual 
YM theory  is defined.

The ansatz for the Janus solution is given by
\begin{eqnarray}
ds^2 &=& f(\mu) \left(d\mu^2 + ds^2_{AdS_{4}}\right)+
      ds_{S^5}^2\ , \nn\\
\phi&=&\phi(\mu)\ , \label{ansatza}\\
F_5&=& 2 f(\mu)^{5\over 2} d\mu \wedge \omega_{AdS_4}
+ 2 \omega_{S^5}\ ,\nn
\end{eqnarray}
where $\omega_{AdS_4}$ and $\omega_{S^5}$ are the unit volume forms
on $AdS_4$ and $S^5$ respectively. Thus, in particular, the five
sphere $S^5$ remains unchanged keeping the $SO(6)$ R-symmetry. But the
supersymmetry is completely broken. The $SO(3,2)$ of $AdS_4$ is clearly
preserved as one can see from the above ansatz.

 The relevant IIB supergravity equations of motion are
given by
\begin{eqnarray}
&&R_{\alpha\beta} -{1\over 2}\partial_\alpha \phi
\partial_\beta \phi
  -{1\over 96} F_{\alpha}^{\phantom{a}\mu\nu\lambda\delta}
F_{\beta\,\mu\nu\lambda\delta}
=0\ ,\nn\\
&&\partial_\alpha(\sqrt{g} g^{\alpha\beta}\partial_\beta \phi)=0
\ ,\label{eqofm}\\
&&*F_5=F_5\ ,\nn
\end{eqnarray}
together with the Bianchi identity $dF_5=0$.
The equation of motion for the dilaton can be integrated
leading to
\begin{equation}
\phi'(\mu) = {c\over f^{3\over 2}(\mu)}\ .
\label{dila}
\end{equation}
The Einstein equations give rise to
\begin{eqnarray}
2 f'f' -2 f f'' &=& -4 f^3 +{c^2\over 2}
    {1\over f}\ ,\nn\\
12 f^2 + f'f' + 2 f f''&=& 16  f^3\,.
\label{einstein}
\end{eqnarray}
It is easy to see that these equations are equivalent to the first
order differential equation
\begin{equation}
f'f' = 4 f^3 -4f^2+{c^2\over 6} {1\over f}\ ,
\label{einsteinb}
\end{equation}
corresponding to the motion of a particle with zero  energy in a
potential given by 
\begin{equation}
V(f) = -4 \left(f^3 -f^2+{c^2\over 24} {1\over f}\right)\ .
\label{potential}
\end{equation}
The solution, $f(\mu,c)$, may be expressed analytically in terms of the
Weierstrass ${\cal P}$-function\cite{D'Hoker}.

\begin{figure}[ht!]
\centering \epsfysize=9cm
\includegraphics[scale=0.7]{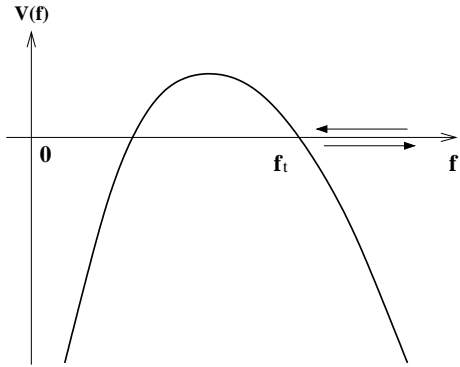}
\caption{\small The dynamics corresponds to the
particle motion under a  potential with zero energy.
The trajectory of our concern is the one in which
the particle starts from infinity, reflected at $f_{t}$ and goes back
to infinity.}
\end{figure}

%
%

Note that $c=0$ corresponds to the unperturbed $AdS_5\times S^5$
with a constant dilaton. For $c\in [0, {9\over 4\sqrt{2}}]$,
the potential has two positive zeros and, as depicted in Figure 1,
we are particularly interested in the trajectory starting from infinity,
 reflected 
and going back to infinity.
Note that $\mu$ is ranged over $[-\mu_0,\mu_0]$ with
\begin{eqnarray}
\mu_0=\int_{f_{t}}^\infty { df \over 2\sqrt{f^3-f^2
+{c^2\over 24}{1\over f}}}\ \ge\  {\pi\over 2}\,\,,
\label{dilbb}
\end{eqnarray}
where 
 $f_{t}$ is the turning point of the potential.
The dilaton in this case varies over a finite range and the string couplings
can be made arbitrarily small;
\begin{eqnarray}
\phi(\mu_0)-\phi(-\mu_0) =  c \int_{-\mu_0}^{\mu_0}
{d\mu \over
    f^{3\over 2}(\mu) }
=c\int_{f_{t}}^\infty { df \over f^{3\over 2} \sqrt{f^3-f^2
+{c^2\over 24}{1\over f}}}\ .
\label{dilb}
\end{eqnarray}

\begin{figure}[ht!]
\centering \epsfysize=9cm
\includegraphics[scale=0.7]{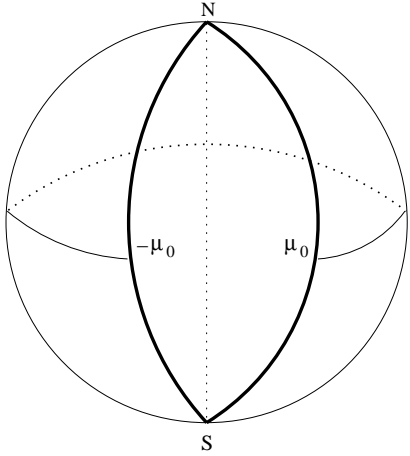}
\caption{\small The conformal diagram of the constant time slice
 is depicted here. Only the surface of $\mu$ and $\lambda$
coordinates is shown, where each point 
corresponds to $S^2$.
}
\label{shape}
\end{figure}

 Adopting the global coordinate for the $AdS_4$
slice
as in (\ref{adssix}), the metric in this case becomes
\begin{equation}
ds^2 = {f(\mu)\over\cos^2\lambda}
 \big( -d\tau^2 + \cos^2\lambda d\mu^2 +
  d\lambda^2 +\sin^2\lambda d\Omega_2^2\big)\ .
\label{geo}
\end{equation}
The spatial section of the conformal metric, i.e. the metric inside
the parenthesis is depicted in Figure \ref{shape}.
Only the surface of 
$\mu$ and $\lambda$ coordinates is
drawn, 
 where each point 
represents $S^2$. 
The boundary consists of two parts; one is at $\mu=-\mu_0$
and the other at $\mu=\mu_0$. These two halves of $S^3$ are joined
through the north and south poles. The dilaton varies from one constant,
$\phi_-=\phi(-\mu_0)$, at one half of the boundary at $\mu=-\mu_0$,
to another, $\phi_+=\phi(\mu_0)$, at the other half of the boundary
at $\mu=\mu_0$, running through the bulk as $\mu$ changes.

In the pure AdS case,
the geometry covers only a half of the surface of the globe with
the range of $\mu \in [-\pi/2, \pi/2]$.
As one  can see from (\ref{dilbb}) for the Janus solution,
the range of $\mu$ becomes larger as
$c$ grows.

\begin{figure}[ht!]
\centering \epsfysize=9cm
\includegraphics[scale=0.7]{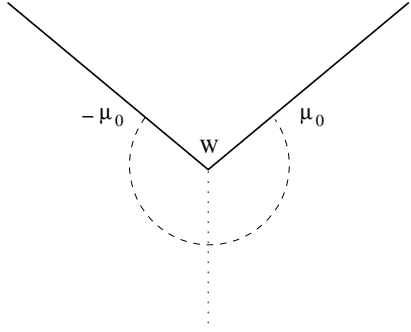}
\caption{\small The conformal diagram of the spatial section 
of the Poincar\`e type coordinate 
is depicted here.  The bulk of the Janus solution
corresponds to the region under the solid line.
Each point on the plane corresponds to ${ R}^2$.
The total boundary again consists of two parts, each of
which has the geometry of
a half of ${R}^3$.  These two parts are joined through $W$, which is
a codimension one interface.
}
\label{shapep}
\end{figure}

When we adopt the Poincar\'e patch for the $AdS_4$ slice as in
 (\ref{adsseven}),
 the metric is written as
\begin{equation}
ds^2 = {f(\mu)\over y^2}
 \big(-dt^2 + d\vec{x}^2 +dy^2 +y^2 d\mu^2
\big)\ .
\label{geoa}
\end{equation}
The conformal mapping of the spatial section is depicted in Figure
\ref{shapep}.
 The bulk of the Janus solution
corresponds to the region under the solid line.
Each point on the plane represents ${ R}^2$.
As in the pure AdS case, $y=\infty$ corresponds to the horizon.
Again the boundary is at $\mu=\pm \mu_0$. Each of these is a half of
${R}^3$, being joined together through the wedge $W$ that is
${ R}^2$.

Again the dilaton varies from one constant, $\phi_-=\phi(-\mu_0)$,
at one half of the boundary $\mu=-\mu_0$, to another,
$\phi_+=\phi(\mu_0)$, at the other half $\mu=\mu_0$,  running
through the bulk as $\mu$ increases.
Hence from the viewpoint of the boundary, the dilaton is constant
in each half, taking the value of $\phi_+$ and $\phi_-$
respectively. We introduce a boundary coordinate $x_3$ by $\pm y$ for $\mu=\pm \mu_0$.
The discontinuity of the dilaton occurs through the joint $W$.
One might worry about possible singularities around the wedge
$W$. However, this is just an artifact of the conformal diagram. 
Nothing singular happens around
the wedge from the viewpoint of the full geometry.

Near the boundary $\mu\to \pm \mu_0$, one can easily  show that the factor
$f$ and the dilaton 
behave as 
\begin{equation}
f= {1\over  (\mu\mp\mu_0)^2}+O(1)\ ,\quad \quad \phi= \phi_\pm \mp
{c\over 4}(\mu\mp\mu_0)^4 + O((\mu\mp \mu_0)^6)\label{expanb}\ .
\end{equation}
In the standard AdS/CFT  dictionary the asymptotic value of the dilaton is
identified with the coupling constant of the four dimensional ${\cal N}$=4 SYM
theory living on the boundary.   The subleading term is identified
with the expectation value for the operator $
\mbox{Tr}(F^2)$. Hence the
interpretation of our solution is that the boundary consists of two
half spaces, given by $\mu= \pm \mu_0$, where the coupling constant is
given by
$g^2=4\pi e^{ \phi_\pm}$ respectively.
This is the holographic dual  suggested in Ref.~\cite{Hirano}.

More precise version of the holographic dual
was identified in Ref. \cite{Karch}. Note first that the boundary
metric of the dual CFT is obtained by multiplying the bulk metric by
$h^2$ where $h$ has a linear zero at the boundary. Different choice
of the factor $h$ leads to the different boundary metric.
The standard dictionary  of AdS/CFT correspondence then says
that the near boundary behaviors of the bulk field, $\phi_\Delta$,
dual to
the operator $O_\Delta$ of dimension $\Delta$ can be expressed in terms
of $h$ by
\begin{equation}
\phi_\Delta \sim  a_h (x) h^{4-\Delta} + b_h (x) h^\Delta  + \cdots
\end{equation}
where $x$ is the boundary coordinate.
Then this corresponds to turning on the source
$\int dx\, a_h (x) O_\Delta (x)$  while the one point function is given by
$\langle O_\Delta (x)\rangle= -(2\Delta - 4) b_h (x)$.
By the choice of $h= y/ \sqrt{f}$, the boundary metric becomes flat
Minkowski $R^{1,3}$.
The operator dual to $\phi$ is the ${\cal N}$=4 SYM Lagrange
density but the precise form is more subtle as noted in Ref.
\cite{Karch}.
The dual operator ${\cal L}'_0$ is the variant of the ${\cal N}$=4 Lagrange
density in which the scalar kinetic term is $X^I \partial^2 X^I/2$
rather than $-\partial X^I \partial X^I/2$.
The action for the Janus dual is given by
\begin{equation}
 S = \int d^4 x {1 \over g^2 (x_3)}\,{\cal L}'_0
\end{equation}
where
\begin{equation}
 {1 \over g^2 (x_3)}= {1\over g_+^2} \Theta (x_3) +
 {1\over g_-^2} \Theta (-x_3) = {1\over \bar{g}^2}
 (1- \gamma \epsilon(x_3))
\end{equation}
with $\gamma = (g_+^2- g_-^2)/(g_+^2+ g_-^2)$ and
$\bar{g}^2= (g_+^2+ g_-^2)/(2g_+^2g_-^2)$.
Here $\Theta(x)$ is the step function and 
$\epsilon(x)$ for the sign function.
The supergravity prediction
with precise coefficient\cite{Balasub,Karch} is given by
\begin{equation}
 e^\phi = e^{\phi_\pm} (1-{2\pi^2\over N^2 }
\langle {\cal L}'
\rangle \, h^4 + \cdots )
\end{equation}
where $ {\cal L}' = {\cal L}'_0/\bar{g}^2 $.
Therefore the expectation value is
\begin{equation}
 \langle {\cal L}'
\rangle =  \epsilon (x_3) {N^2\over 2\pi^2}
{c\over 4 x_3^4}\,.
\end{equation}

Then one can confirm the above gravity prediction on the
one point function using the conformal perturbation theory
in the gauge theory side\cite{Karch}.

Finally if $c$ exceeds $c_{cr}=9/(4\sqrt{2})$, the zero energy motion of
particle reaches the point $f=0$, where the geometry develops a
naked singularity. The singularity is timelike here. As one
approaches
the singularity, the magnitude of dilaton diverges. Hence the string
theory becomes either extremely weakly coupled or strongly coupled.
The dual field theory description is not identified in this regime.
But we note that the diverging dilaton is to do with the appearance
of the curvature singularity. Since the geometry involves the timelike
naked
singularity, whether this regime is physical or not is unclear.

\section{Meaning of the bulk geometry}

The bulk in the AdS/CFT correspondence may be
interpreted as
the space of  RG scale in CFT side\cite{Verlinde}. The gravity
on-shell action is identified with the generating functional
of the connected graphs of the boundary quantum field theory.
The correlation functions computed this way exhibit divergences
due to the infinite volume and this is the place where the
renormalization becomes relevant.
The flow into the bulk direction may  be
interpreted as the RG flows as we change the cut-off scale
of the dual gauge theory.
By the Legendre transform of the generating functional,
one gets the quantum effective action defined at a
given cut-off scale.

In the Janus gauge theory, the YM coupling abruptly changes
from $g_+$ to $g_-$, which is encoded at the boundary values
$\phi(\pm\mu_0)$.
For the definition of the  boundary field theory,
one needs only information of $\phi(\mu)$ at the boundary since the
boundary field theory is defined at infinite cut-off scale
in the momentum space.

What is the meaning of the precise form of
$\phi$ as a function of $\mu$ in the Janusian case?
We would like argue below that this bulk running
of the dilaton can have its meaning if one lowers
the cut-off scale  to a finite value.

Following the prescription of Ref.~\cite{Verlinde}, the dilaton
at a finite cut-off here may be interpreted as a renormalized coupling 
at a given RG scale. In order to describe the details of the 
RG flow, the Fefferman-Graham coordinate, as adopted in Ref.~\cite{Verlinde}, 
is most convenient. Unfortunately,
for the case of the Janus geometry, the Fefferman-Graham
coordinate does not cover the whole region of the bulk\cite{Papadimitriou}. 
Rather it covers only the region of  $\mu_L\le
\mu\le \mu_0$ with nonvanishing $\mu_L$ excluding some bulk part
around the wedge of Figure \ref{shapep}.
Thus in this coordinate, the study of renormalization
at finite cut-off scale is problematic. 
For this reason, to carry out the holographic renormalization,
one has to develop a new technique with generic lapse and shift
functions. This is certainly interesting direction
for a further study but beyond the scope of 
this note. For our present purpose, we shall work in a particular 
metric that can cover the whole of the Janus geometry:
\begin{equation}
 ds^2= N^2 d(\Lambda^{-1})^2 +
g_{ij} (dx^i + N^i d(\Lambda^{-1}))  (dx^j + N^j d(\Lambda^{-1}))\,,
\label{lapse}
\end{equation}
where we take 
\begin{equation}
g_{ij} dx^i dx^j
=\Lambda^2 (-dt^2 + dx_1^2 +dx_2^2 +dx_3^2)
\label{lapse1}
\end{equation}
with an appropriate choice of $\Lambda$, $x_3$, $N$ and $N^i$.
Since $\Lambda$ is responsible for the scaling of $g_{ij}$, it can be 
interpreted as a physical cut-off scale 
coordinate\cite{Verlinde}.

Comparing (\ref{lapse}), (\ref{lapse1}), and (\ref{geoa}),
one finds that the cut-off scale coordinate
$\Lambda$ is  $\sqrt{f(\mu)}/y$. Then along constant $\Lambda$,
$x_3$ coordinate is given by
\begin{equation}
 dx_3^2= dy^2 + y^2 d\mu^2 = \left( f^2 + {c^2\over 24f^2}\right)
 {d\mu^2\over \Lambda^2}
\end{equation}
such that
\begin{equation}
 ds^2 = \Lambda^2 (-dt^2 + dx_1^2 +dx_2^2 +dx_3^2)\,,
\label{cft}
\end{equation}
for constant $\Lambda$ slice.
Explicitly $x_3$ reads 
\begin{equation}
x_3= {1\over \Lambda}\int^\mu_0 d\mu
\sqrt{ f^2 + {c^2\over 24f^2}} \equiv {1\over \Lambda} G(\mu)\,.
\end{equation}

For $c=0$, $f_0 = 1/\cos^2\mu$ and $1/\Lambda= y \cos\mu$. $G(\mu)$ can
be obtained as $G=
\tan \mu$ and, then, $x_3=y\sin\mu$. The full metric in this case becomes
the standard form of the Poincar\`e metric,
\begin{equation}
 ds^2 = \Lambda^2 (-dt^2 + dx_1^2 +dx_2^2 +dx_3^2+ (d\Lambda^{-1})^2)\,.
\end{equation}

One can express $\phi(\mu)$  as a function of $\Lambda$ and $x_3$ by
\begin{equation}
\phi(\mu)= \phi( G^{-1} (\Lambda x_3)).
\end{equation}
We interpret this $e^{\phi(x_3)}$ as the $x_3$ dependent coupling squared
seen at the scale $\Lambda$, which becomes now a smooth
function of $x_3$ at any finite $\Lambda$.

\section{Cosmological Solutions}

In this section we like to discuss about the cosmological solution
as a dilatonic deformation of $AdS_5$ geometry. The solution involves
big-bang and big-crunch singularities
and how to interpret these in terms of dual
field theory is our main concern.

The FRW type  cosmological
solution  can be obtained from the Janus solution by
performing the double analytic continuation\footnote{This is not 
the conventional 
Wick rotation of quantum field theories. For instance, the ranges of $\mu$
and $\alpha$ below differ from each other. Here it is rather a formal way of 
obtaining new solutions and field theories.};
\begin{equation}
X_0 \rightarrow -i X_{d-1}\,, \ \ \   X_{d-1} \rightarrow  i  X_0
\label{double}
\end{equation}

We note that the ansatz in (\ref{ansatza}) for the Janus solution
can be rewritten as
\begin{eqnarray}
ds^2 &=& {f(w)\over 1+w^2} ds^2_{AdS_{5}} +
      ds_{S^5}^2\ , \nn\\
\phi&=&\phi(w)\ , \label{ansatzaa}\\
F_5&=& 2 \left({f(w)\over 1+w^2}\right)^{5\over 2} \omega_{AdS_5}
+ 2 \omega_{S^5}\ ,\nn
\end{eqnarray}
where $\omega_{AdS_5}$ is the unit volume form
on $AdS_5$.
Then by 
the above double analytic continuation, the ansatz becomes
\begin{eqnarray}
ds^2 &=& {f(X_0)\over 1-X_0^2} ds^2_{AdS_{5}} +
      ds_{S^5}^2\ , \nn\\
\phi&=&\phi(X_0)\ , \label{ansatza1}\\
F_5&=& 2 \left({f(X_0)\over 1-X_0^2}\right)^{5\over 2} \omega_{AdS_5}
+ 2 \omega_{S^5}\ ,\nn
\end{eqnarray}
which is valid only for $|X_0| \le 1$. With the transformation
of
\begin{eqnarray}
c \rightarrow  -ic\ ,
\end{eqnarray}
the dilaton equation is solved by
\begin{equation}
(1-X_0^2)\phi'(X_0) = {c\over f^{3\over 2}(X_0)}\ .
\label{dila1}
\end{equation}
Also the Einstein equations are reduced to
\begin{equation}
(1-X_0^2)^2 (f'(X_0))^2 = -4 f^3 +4f^2 +{c^2\over 6} {1\over f}\ .
\label{einsteinb1}
\end{equation}

Before going into details, let us briefly comment on the case
of $X_0^2 > 1$. This case  cannot be obtained
from the Janus
solution
by the
double analytic continuation.  But one can  compute equations of motion
directly starting from
the ansatz,
\begin{eqnarray}
ds^2 &=& {f(X_0)\over X_0^2-1} ds^2_{AdS_{5}} +
      ds_{S^5}^2\ , \nn\\
\phi&=&\phi(X_0)\ , \label{ansatza2}\\
F_5&=& 2 \left({f(X_0)\over X_0^2-1}\right)^{5\over 2} \omega_{AdS_5}
+ 2 \omega_{S^5}\ .\nn
\end{eqnarray}
Then
the type IIB  equations of motion are reduced to 
\begin{eqnarray}
(X_0^2-1) \phi'(X_0)&=&{c \over f^{3/2}}\ ,\\
(X_0^2-1)^2(f'(X_0))^2&=&4f^2+ 4f^3
+{c^2\over 6}f^{-1}\ .
\end{eqnarray}
There is an extra change in the sign of $f^3$ term,
which is due to the change of signs
in the term of five form squared. 

Since the potential is negative definite, the particle
hits the $f=0$ region inevitably.  From the form of the
scalar curvature,
\begin{equation}
R=-20-{c^2\over 2f^4}\ ,
\label{scurv}
\end{equation}
one can see that
the solution involves curvature singularities.
The singularity is timelike as noted in Ref. \cite{Hirano}
and nothing to do with the cosmological singularities.
For this reason, we shall not discuss this branch of solution
further here.

\begin{figure}[ht!]
\centering \epsfysize=9cm
\includegraphics[scale=0.7]{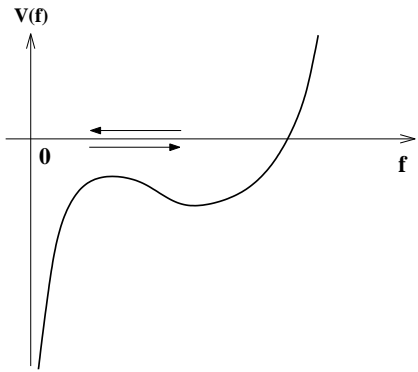}
\caption{\small The dynamics again corresponds to the
particle motion under a  potential with zero energy. The particle
starts from zero, reflected at the turning point  and goes back
to zero.}
\label{cos}
\end{figure}

Now let us get back to the case of $|X_0|<1$.
By the change of variable $X_0= \tanh \alpha$,
the equation (\ref{einsteinb1}) becomes
\begin{equation}
\left({df\over d\alpha}\right)^2 = -4 f^3 +4f^2 +{c^2\over 6} {1\over f}\ ,
\label{einsteinb11}
\end{equation}
which can be again interpreted as a particle motion with a zero
energy under a potential,
\begin{equation}
V(f) = 4 \left(f^3 -f^2-{c^2\over 24} {1\over f}\right)\ .
\label{potential1}
\end{equation}
The shape of the potential is depicted in Figure \ref{cos}.

For $c=0$, the solution corresponds to the unperturbed $AdS_5$ with
$f= 1-X_0^2=1/\cosh^2 \alpha$. The range of $\alpha$ is given by
$\alpha\in [-\infty,\infty]$.

If we turn on nonvanishing $c$, the range of $\alpha$
becomes finite now. By
choosing $\alpha=0$ at the turning point, the range for $\alpha$ is given
by $[-\alpha_0, \alpha_0 ]$ where $f(\pm \alpha_0)=0$.
For small $f$ region, $V \sim  - c^2/(6f)$ and the equation is
approximated as
\begin{equation}
{df\over d\alpha} \sim \pm {|c|\over \sqrt{6 f}}\ ,
\end{equation}
which is solved by
\begin{equation}
f^{3\over 2}  \sim  {3 |c|\over 2\sqrt{6}} |\alpha\pm \alpha_0|\ .
\end{equation}

Then the dilaton behaves as
\begin{equation}
\phi  \sim \pm  {2\sqrt{6}\over 3}\ln |\alpha\pm \alpha_0|\ .
\end{equation}
Hence for positive/negative  $c$,
the dilaton starts from $\mp\infty$, monotonically
increases/decreases and ends up with $\pm\infty$.

The scalar curvature in (\ref{scurv}) become singular
as $\alpha\rightarrow \pm \alpha_0$. These singularities
are spacelike cosmological singularities.
The precise form of the solution, $f$,  can be obtained from the Janus solution,
$f_J (\mu, c)$, by
\begin{equation}
f(\alpha , c) = f_J( i\alpha, -ic)\,.
\end{equation}

\begin{figure}[ht!]
\centering \epsfysize=9cm
\includegraphics[scale=0.7]{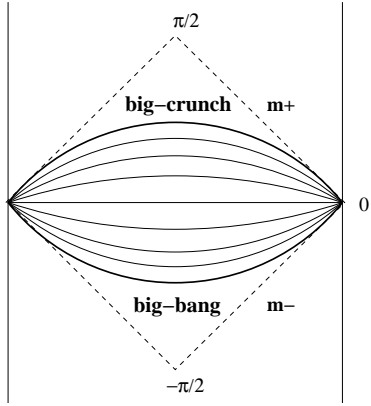}
\caption{\small The Penrose diagram is depicted here for the
dilatonic cosmological solution. The two parallel lines
represent the boundary of original global $AdS_5$ space. The upper thick solid 
curve corresponds to the big-crunch whereas the lower one 
to the big bang singularity.}
\label{pen}
\end{figure}

Introducing the new coordinate $\tau$ by $\sin \tau= \tanh \alpha$,
 the five dimensional metric can be presented in the FRW form,
\begin{equation}
ds^2= {f(\tau)\over \cos^2\tau}( -d \tau^2+ \cos^2 \tau ds^2_{EAdS_4}) \ ,
\label{glo2}
\end{equation}
where $ds^2_{EAdS_4}$ is the metric of the Euclidean $AdS_4$
space,
\begin{equation}
ds^2_{EAdS_4} = {dr^2\over 1+r^2} + r^2 ds^2_{S^3}=
d\eta^2+ \sinh^2\eta \, ds^2_{S^3}\ .
\end{equation}
Thus the solution respects $SO(4,1)\times SO(6)$ symmetries.
The Penrose diagram for this cosmology is depicted in Figure \ref{pen}.

If one uses the Poincar\'e patch coordinate for the Euclidean
$AdS_4$, the metric can take the form,
\begin{equation}
ds^2= {f(\alpha)\over y^2}( -y^2d\alpha^2  +
dy^2 + dx^2_1+ dx_2^2+ dx_E^2 ) \ .
\label{poin2}
\end{equation}

Unlike the case of the usual Poincar\'e patch and global coordinates of
$AdS_5\,\,$\cite{Aharony},  the Penrose diagram for (\ref{poin2})
is the same as the one for the spacetime (\ref{glo2}). Let us illustrate
how this happens. We are interested in the conformal structure of 
$(y,\alpha)$ directions, so a point in the Penrose diagram corresponds
to a plane spanned by $(x_1, x_2, x_E)$. The conformal structure
can be obtained from 
\begin{equation}
 -y^2d\alpha^2  +
dy^2 = -dt^2 + dy^2 \,,
\end{equation}
with $t=y \sinh\alpha$ and $x=y \cosh\alpha$. 
The coordinates $\alpha$ and $y$ are respectively  ranged over
$[-\alpha_0, \alpha_0]$ and $[0,\infty)$ and $\alpha=\pm\alpha_0$
are the locations of the spacelike singularities.
If $\alpha$ were ranged over $(-\infty, \infty)$, then 
$(y,\alpha)$ space covers the whole quadrant specified by
$x+t \ge 0$ and $x-t \ge 0$. The singularities are then obviously residing 
within the quadrant.  We now conformally compactify the space $(t, x)$
by
\begin{equation}
-dt^2 + dy^2 = - \sec^2 w_+ \sec^2 w_-  dw_+ dw_-  \,,
\end{equation}
where $\tan w_\pm = t \pm x$ with $w_\pm \in [-\pi/2,\pi/2]$.
The boundary of the usual Poincar\'e patch is at $w_+ - w_-=0$
corresponding to the right vertical line of  Figure 5.
The quadrant in the above is now covered by $0\le w_+\le \pi/2$
and $-\pi/2 \le w_- \le 0$. By carefully identifying  locations
of the singularities of $\alpha=\pm \alpha_0$ within the quadrant, 
one finds the Penrose diagram of Figure 5.

\section{Dual of  Big-bang and  Big-crunch}

Our proposal for the dual of the cosmological solution is
as follows. Since the cosmological solution can be obtained
by the double analytic continuation described in the previous section,
we propose that the corresponding dual gauge field theory
can be again defined by the procedure of the double analytic continuation. 
But the coupling of the dual field theory is time dependent
and this may interfere with the  process of the
renormalization. Because of this, the field theory may
not be defined at infinite cut-off scale.

Thus we do not perform the double analytic continuation 
directly in the Janus
dual gauge theory defined at the boundary. Instead, we note that there is a
generating functional for the renormalized correlation functions at
a given RG scale. Or alternatively one may talk about the quantum
effective action by the Legendre transformation. We perform the double 
analytic continuation at the level of the generating functional.
If one has full nonperturbative knowledge of the generating
functional including all order $1/N$ corrections, the procedure
of the double analytic continuation can be made precise leading to the
exact gauge theory dual of the cosmology.
As we will see more
 below, the double analytic continuation for the gauge
theory on (\ref{cft}) is
given by
\begin{eqnarray}
&&t \rightarrow \ -ix_E, \ \ \ \ x_3(\mu, c) \rightarrow
 i x_0=x_3(i\alpha, -ic)\nn\\
&& \mu \rightarrow  \ \ i\alpha, \ \ \ \ \ \ \ c\rightarrow  -i c\ ,
\end{eqnarray}
in the leading order supergravity approximation.

If one ignores the issue of renormalization, the quantum effective action
corresponds roughly
to the ${\cal N}$=4 SYM Lagrangian density with time dependent
YM coupling. At the big-bang or the big-crunch then, the YM coupling goes
either to
zero or to infinity. For both cases, the supergravity approximation
breaks down. Namely for zero coupling limit, the curvature radius
$l = (g^2 N)^{1\over 4}\, l_s$ becomes zero and
the $\alpha'$ expansion breaks down completely. For the strong coupling
limit, the string tree level approximation breaks down completely too.
This way we may understand the appearance of the cosmological
singularity, which signals simply the breaking down of the
geometric description. We further note that the strong coupling limit is
related to the weak coupling limit by the S-duality
symmetry of the type IIB superstring theory.
 Thus one may see that in this correspondence,
the big-bang and big crunch are related by the S-duality of
the ${\cal N}$=4 SYM theory. Since the gauge theory is time dependent 
now, the in- and the out-vacuum will differ and there is 
the effect of particle production in general. Hence 
the S-duality does not mean that the in- and the out-vacuum are the same.

To see the indication of break down of the geometric description,
we use the coordinate (\ref{poin2}). The scale is then again
identified with $\Lambda =f^{1\over 2}/y$. For the constant
$\Lambda$, we defined $x_0$ coordinate by
\begin{equation}
 -dx_0^2= dy^2 -y^2 d\alpha^2 = -\left( f^2 - {c^2\over 24f^2}\right)
 {d\alpha^2\over \Lambda^2}
 \label{x0}
\end{equation}
such that
\begin{equation}
 ds^2 = \Lambda^2 (-dx_0^2 + dx_1^2 +dx_2^2 +dx_E^2)\,,
\label{cft1}
\end{equation}
for constant $\Lambda$ surface.
Then $x_0$ is given by
\begin{equation}
x_0= {1\over \Lambda}\int^\alpha_0 d\alpha
\sqrt{ f^2 - {c^2\over 24f^2}} \equiv {1\over \Lambda} K(\alpha)\,.
\end{equation}

For $c=0$, $f_0 = 1/\cosh^2 \alpha$ and $1/\Lambda= y \cosh \alpha$.
The function $K$ can
be obtained as $K=
\tanh \alpha$ and, then, $x_0=y\sinh \alpha$. The full metric for $c=0$
becomes
the standard form of the Poincar\`e metric,
\begin{equation}
 ds^2 = \Lambda^2 (-dx_0^2 + dx_1^2 +dx_2^2 +dx_3^2+ (d\Lambda^{-1})^2)\,.
\end{equation}

Note that 
one major difference from the Janusian case is that the boundary at
infinity
is no longer available
meaning that it is now just a point in the Penrose diagram of
Figure 5.

For non zero  $c$, the definition of $x_0$ breaks down at
$\alpha=\pm\alpha_H$
defined by
\begin{equation}
f(\pm\alpha_H)= {\sqrt{|c|}\over(24)^{1\over 4}}
\end{equation}
due to this signature change of
the right hand side of (\ref{x0}).
For $\alpha_0> |\alpha|> \alpha_H$, the double analytic continuation also
breaks down due to the signature problem.  

Whether one can avoid this signature flip by the other choice
of coordinate in the holographic renormalization is
 not entirely clear. Due to the presence of the 
singularities and the behavior of dilaton, the geometrical
description will be breaking down in any ways and the dual gauge theory 
description can only be defined by the double analytic continuation
of the correlators of the Janus gauge theory.

Since our proposed correspondence is not the direct relation between
the geometry and  the boundary gauge theory
via the AdS/CFT dictionary,
an intuition from geometry is limited anyway and 
we shall not further clarify  meaning of the break-down in the above.

However, since in the weak
coupling limit,
the ${\cal N}$=4 SYM perturbation theory is well defined clearly, we see that
only the geometric description becomes problematic.
We know clearly that the {\sl classical} ${\cal N}$=4 YM theory
in the weak coupling limit
 is nothing like the ten  
dimensional supergravity. In this respect, the failure of description
in this regime is rather obvious.
Therefore, 
the time dependent field
theory is well defined for all $x_0$ and the appearance of the
cosmological singularity is simply due to the failure of
supergravity description.

\begin{figure}[ht!]
\centering \epsfysize=6cm
\includegraphics[scale=0.7]{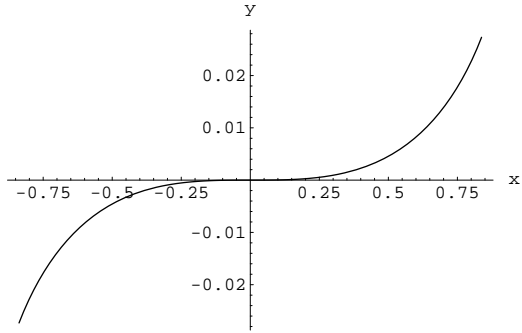}
\caption{\small The behavior of the dilaton as a function of $\Lambda x_0$
is drawn here for $c=0.1$. The x-axis is for $\Lambda x_0$ and 
the y-axis for $\phi(\Lambda x_0)-\phi(0)$.
One finds that the maximum value of
$|x_0\Lambda|$ is numerically 0.907 at the sign flip and 
the value of $|\phi(\lambda x_0)-\phi(0)|$ there 
is 0.0423.}
\label{dil}
\end{figure}

Finally, with help of the function $K$,
$\phi(\alpha)$ can be expressed  as a function of $\Lambda$ and $x_0$ by
\begin{equation}
\phi(\alpha)= \phi( K^{-1} (\Lambda x_0)).
\label{scale}
\end{equation}
Then $e^{\phi(\Lambda x_0))}$ can be interpreted as the $x_0$
dependent YM coupling squared
seen at scale $\Lambda$. We illustrate $\phi(\Lambda x_0)$
in Figure \ref{dil} for $c=0.1$. 
As a function of $x_0$, the shape 
of $\phi$ simply scales as $\Lambda$ varies,  which
 is obvious from (\ref{scale}).
The slope becomes steeper as $\Lambda$ becomes larger.

\section{Discussion}

In this note, we obtain the dual  gauge field theory of cosmological
solution by the procedure of double analytic continuation starting from
the Janus solution and its gauge theory dual. We note that
the big-bang and big-crunch singularities in this
cosmology are resolved naturally in the dual gauge theory
description. These cosmological singularities signal
simply a failure of the supergravity
description
of the full IIB string theory.
Our analysis of the double analytic continuation is based on the
supergravity approximation and there might be a chance to
improve the analysis if part of the supersymmetries are preserved.
Recently supersymmetric version of Janus
solution\cite{Karch,Clark,D'Hoker,D'Hoker1} has been
investigated and this might be the place where one has
a better control of corrections.

The conformal compactification of Janus solution leading to the
boundary gauge theory on $R\times S^3$ can be discussed similarly and we
have not presented details of the computation.
Note that there is yet another compactification of the Janus solution
leading  another boundary conformal field theory. Namely
the conformal factor $h$ can be chosen as $1/\sqrt{f}$.
For this case as noted in \cite{Karch}, the boundary geometry
is given by two $AdS_4$ joined at their boundary
$R\times S^2$. The change of cut-off scale then corresponds to
the change of $\mu$ as $f$ is only function of $\mu$.
 As we are away from the boundary by changing $f(\mu)$ $(= f(-\mu) )$
down to a
finite value, the two $AdS_4$ spaces at $\mu$ and $-\mu$
are joined at their boundary.
Since the  coupling is only function of the scale $\mu$, the coupling
remains constant
for each $AdS_4$ while there is a jump through the common boundary.
This is contrasted the case
of
the choice of
$h=y/\sqrt{f}$, where
the
coupling shows again step function dependence of $x_3$ at the UV limit.  But,
when we lower the cut-off scale  to a finite value, the coupling
varies smoothly
as a function of $x_3$.

For the above $AdS$ type compactification, what derives the change of coupling
from the view point of dual gauge theory is not quite clear.

We now consider performing the double analytic continuation of
this new compactification. The scale coordinate $\mu$ turns into
the time coordinate. The $AdS_4$ becomes Euclidean
$AdS_4$ by the analytic continuation. Thus the scale change turns into
the cosmological time flow. We are dealing with set of Euclidean
$AdS_4$ with changing scale and coupling along the time flows.
The failure of description
and the curvature singularity is again due to the weak and
strong coupling limit, by which the supergravity description
breaks down.
But how to understand  such a collection of field theories arranged
by the
cut-off scale and couplings is not
clear to us.

Finally, there is the issue of  how the holography\cite{tH,Susskind}
is realized in the cosmological context\cite{Bak2,Bousso}.
The above cosmology/gauge theory correspondence may serve as a solid
framework for the discussion of cosmological holography.
In particular the time dependence of  the cut-off scale seems
important in understanding the number of
degrees of the dual gauge theory. Further studies are required in
this direction.

\section*{Acknowledgments}
We are grateful to Kimyeong  Lee for useful discussions and
conversations.
This work  is supported in part by
 KOSEF ABRL R14-2003-012-01002-0, 
KOSEF R01-2003-000-10319-0, KOSEF SRC CQUeST R11-2005-021,
and KRF-2003-070-C00011.

\end{document}